\documentclass[reprint,aps,pre,amsmath,amssymb,amsthm,superscriptaddress,showpacs]{revtex4-1}
\usepackage{graphicx}
\usepackage{hyperref}

\newtheorem{definition}{Definition}

\begin{document}

\title{Percolation in higher order networks via mapping to chygraphs}

\author{Alexei Vazquez}
\email{alexei@nodeslinks.com}
\affiliation{Nodes \& Links Ltd, Salisbury House, Station Road, Cambridge, CB1 2LA, UK}

\begin{abstract}
Percolation theory investigates systems of interconnected units, their resilience to damage and their propensity to propagation. For random networks we can solve the percolation problems analytically using the generating function formalism. Yet, with the introduction of higher order networks, the generating function calculations are becoming difficult to perform and harder to validate. Here, I illustrate the mapping of percolation in higher order networks to percolation in chygraphs. Chygraphs are defined as a set of complexes where complexes are hypergraphs with vertex sets in the set of complexes. In a previous work I reported the generating function formalism to percolation in chygraphs and obtained an analytical equation for the order parameter. Taking advantage of this result, I recapitulate analytical results for percolation problems in higher order networks and report extensions to more complex scenarios using symbolic calculations. The code for symbolic calculations can be found at \href{https://github.com/av2atgh/chygraph}{github.com/av2atgh/chygraph}.
\end{abstract}

\maketitle

\section{Introduction}

The field of network science has matured and diversified. In recent years there has been an explosion of work on higher order networks, including multiplex networks, interacting networks, interdependent networks, their extensions to hypergraphs and simplicial complexes \cite{kivela14, battiston20, battiston21,bianconi21}. These mathematical constructions capture more elements of the real systems they represent. 

Calculating the giant component is a fundamental problem in network science. For networks ensembles we can solve the problem with  the generating function formalism \cite{molloy98,callaway00}. The generating function formalism has been extended to hypergraphs \cite{ghoshal09}, clique/simplicial-complexes percolation \cite{coutinho20,zhao22,xu22,bianconi23}, multi-type/multi-layer networks \cite{vazquez06a,allard09}, multiplex hypergraphs \cite{sun21}, interacting graphs and hypergraphs \cite{leicht09,bianconi17} and interdependent networks \cite{buldyrev10,radicchi15}. However, with the increase of the higher order complexity, there is a significant growth in the number of independent calculations that are hard to grasp by any individual researcher. Should the field continue in that direction its risks to become fragmented.

To consolidate the work on higher order networks, I introduced complex hypergraphs (chygraphs), a combinatorial construction defined as a set of complexes where complexes are hypergraphs with vertex sets in the set of complexes \cite{vazquez22}. My goal was to apply results on chygraphs to all higher order networks in its class. In particular, I have obtained an analytical solution to the emergence of a giant component on chygraphs \cite{vazquez22}. With this result at hand, solving percolation problems in higher order networks is reduced to plugging in the parameters specific to the structure under consideration and getting the result via symbolic calculations.

Here I illustrate the mapping of higher order networks to chygraphs to solve percolation problems. The report is organized as follows. In Sec. \ref{sec_chygraph} I reintroduce the definition of chygraphs \cite{vazquez22} and illustrate the mapping of graphs, hypergraphs and higher order networks to chygraphs. In Sec. \ref{sec_order_parameter} I go over the main result of the generating function calculation in Ref. \cite{vazquez22}, an analytical expression for the order parameter of the phase transition from disconnected clusters to the emergence of a giant component. For the first time, I report the function to perform the symbolic calculation of the order parameter. Then, in Secs. \ref{sec_graphs_hypergraphs}-\ref{sec_interacting} I illustrate the application of the symbolic calculation to specific examples. Finally, I end with the conclusions in Sec. \ref{sec_conclusions}.

\section{Chygraphs}
\label{sec_chygraph}

Combinatorial constructions help to abstract complex systems into their basic units and their interactions. A {\em graph} or network $G(V, E)$ is a set vertices $V$ and a set of edges $E$, where edges are pairs of vertices. The vertices (nodes) represent the system basic units. The edges (links) represent the interactions between nodes. In some systems the associations between the basic units goes beyond pairwise interactions. A {\em hypergraph} $H(V, E)$ is a set vertices $V$ and a set of hyperedges $E$, where hyperedges are subsets of $V$ ($E=\{e_i\subset V, i=1,\ldots,m\}$). These constructions can be extended to higher order networks by taking into account the existence of layers with different properties and/or interacting between them \cite{kivela14, battiston20, battiston21}. My key point here is that many of these higher order network definitions can be encapsulated in the following combinatorial construction \cite{vazquez22}

\begin{definition}
A complex hypergraph (chygraph) is a set of complexes where complexes are hypergraphs with a vertex set in the set of complexes \cite{vazquez22}.
\label{chygraph}
\end{definition}

This concise definition spells two key properties of complex systems: self-reference and fine-grained structure. Self-reference: Complexes are build from other complexes and they are the building blocks for other complexes as well. Fine-grained structure: The complexes within a complex are organized as a hypergraph.

The complexes are characterized by different properties. We say a complex A includes a complex B when B is in the complex set of A. In turn, we say B is included in A.  The {\em chy-degree} of a complex A is the number of other complexes including A. The {\em cardinality} of a complex is the number of complexes it includes. An {\em atom} is a complex that does not includes any complexes. A chygraph where the complexes are hyperedges (hypergraphs with one hyperedge) is called an {\em ubergraph}, as originally defined by Joslyn and Nowak\cite{joslyn17}. Finally, a chygraph may contain {\em layers} setting apart different types of complexes.

\begin{figure}[t]
\includegraphics[width=3in]{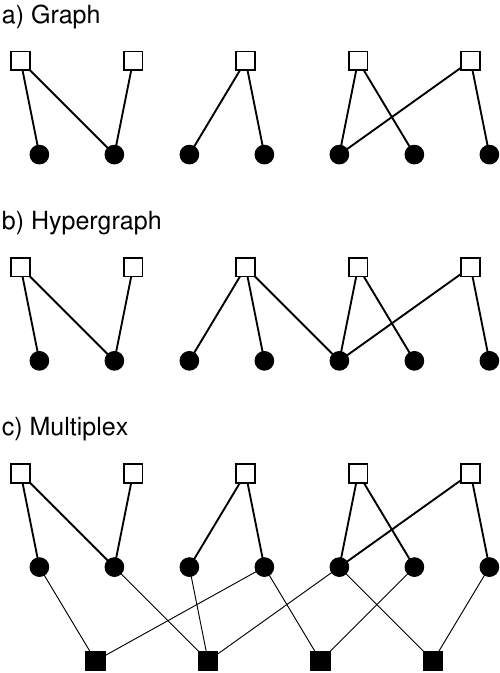}
\caption{Chygraph representation of a graph, hypergraph and multiplex hypergraph. Nodes are represented by circles and edges/hyperedges by squares.}
\label{fig_mappings}
\end{figure}

\begin{figure}[t]
\includegraphics[width=3in]{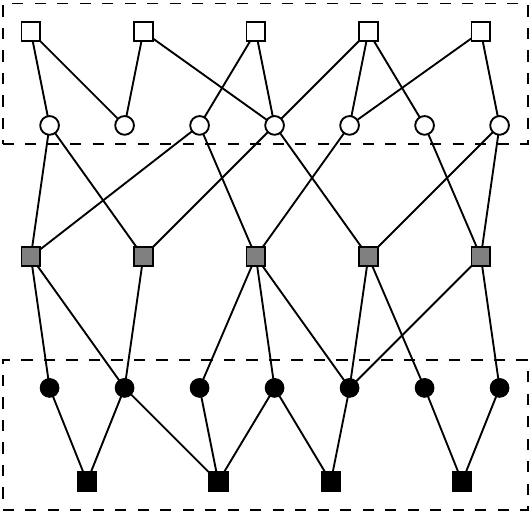}
\caption{Chygraph representation of two interacting  hypergraphs. Nodes are represented by circles and hyperedges by squares. One hypergraph is colored black and the other white. The hyperedges containing nodes in both hypergraphs are colored gray.}
\label{fig_interacting}
\end{figure}

Let us unravel these definitions with some examples. A graph is represented by a two layer chygraph (Fig. \ref{fig_mappings}a). One layer of atoms representing the graph nodes and another for the graph links. Nodes are included in any number of links and links include exactly two nodes. A hypergraph is similar to a graph where links (hyperedges) include any number of nodes (Fig. \ref{fig_mappings}b). A multiplex graph is a chygraph with a layer of atoms representing the nodes and layers of complexes corresponding to different link types (Fig. \ref{fig_mappings}c). Two interacting hypergraphs are represented by  two hypergraphs mappend to chygraphs, plus an additional layer of complexes representing the hyperedges containing nodes from both hypergraphs (Fig. \ref{fig_interacting}). This construction can be extended to any number of hypergraphs introducing one additional layer for each pair of interacting hypergraphs.

\section{Order parameter of the percolation transition}
\label{sec_order_parameter}

Given a complex we can reach the complexes it includes and the complexes it is included in. Repeating this process recursively we will reach a subset of the set of complexes, which in some instances may be the whole set of complexes. That subset of complexes is called a {\em component}. The size of a component is the number of complexes it contains. A giant component is a component with a size of the order to the number of complexes.

A chygraph with several complexes and a few complex-to-complex inclusions has many components of small size. As the number of complex-to-complex inclusions increases there will be a point where a giant component will emerge. Using the generating function formalism I have characterized the transition from finite size components to the emergence of a giant component \cite{vazquez22}. The central object is the 6 dimensional tensor
\begin{eqnarray}
(A_{--})^{ml}_{nk} &=& \delta_{mn}\delta_{lk} - \langle\kappa\rangle_{nk} \delta_{nl},
\nonumber\\
(A_{-+})^{ml}_{nk} &=&  - \left[\langle\bar{s}\rangle_{nk}\delta_{mk} 
+ \langle s\rangle_{nk} (1-\delta_{mk}) \right]
\delta_{nl},
\nonumber\\
(A_{+-})^{ml}_{nk} &=& - \left[\langle\bar{\kappa}\rangle_{nk}\delta_{mk} 
+ \langle\kappa\rangle_{nk}(1-\delta_{mk})\right]
 \delta_{nl},
\nonumber\\
(A_{++})^{ml}_{nk} &=& \delta_{mn}\delta_{lk} - \langle s\rangle_{nk} \delta_{nl},
\label{A}
\end{eqnarray}
where $\delta_{ij}$ is the Kronecker delta ($\delta_{ij}=1$ if $i=j$ and 0 otherwise) and the upper bar denotes the excess averages
\begin{equation}
\langle\bar{\kappa}\rangle_{nk}  =\frac{\langle\kappa(\kappa-1)\rangle_{nk}}
{\langle\kappa\rangle_{nk}},
\label{kappa_excess}
\end{equation}
\begin{equation}
\langle\bar{s}\rangle_{nk}  = \frac{\langle s(s-1)\rangle_{nk}}
{\langle s\rangle_{nk}}.
\label{s_excess}
\end{equation}
The interpretation of $A$ is as follows. The indexes $m$, $l$ and $k$ label layers from $0$ to $L-1$, where $L$ is the number of layers. All terms contain the factor $\delta_{nl}$, indicating that we are dealing with 3 layers: coming from $m$, arriving at $l$ and going to $k$. The equations defining $(A_{ij})^{ml}_{lk}$ represent the self-consistency between the expected component size at layer $l$ coming from layer $m$, with the expected component size at layer $k$ coming from layer $l$. $\delta_{mn}\delta_{lk}$ stands for the component size at layer $l$ coming from $m$.  Excluding the latter, $-(A_{ij})^{ml}_{lk}$ is the expected component size that can be reached coming from a complex at layer $m$ that is included ($i=-$) or includes ($i=+$) a complex C in layer $l$ and then recursively following the complexes that C includes from layer $k$ ($j=-$) and the complexes in layer $k$ that include C ($j=+$).

The matrix $\langle\kappa\rangle_{nk}$ is the expected number of inclusions of a given complex in layer $n$ into complexes in layer $k$. The matrix $\langle s\rangle_{nk}$ is the expected number of inclusions of complexes in layer $k$ by a complex in layer $n$. The matrix $\langle\bar{\kappa}\rangle_{nk}$ is the excess chy-degree. When $m=k$, there is a bias proportional to the chy-degree of layer $n$ complexes to layer $m$ and the inclusion used to come from layer $m$ into $n$ is subtracted when coming back to layer $m$. $\langle\bar{s}\rangle_{nk}$ is the excess inclusion size. When $m=k$, there is a bias proportional to the number of inclusions of layer $n$ complexes into layer $m$ complexes and the inclusion used to come from layer $m$ into $n$ is subtracted when coming back to layer $m$.

The order parameter $\Lambda$ is the maximum eigenvalue of the matrix $-{\rm vec}^2A$
\begin{equation}
\Lambda = \max_{\lambda \in  {\rm eigenvals}(-{\rm vec}^2A)}\lambda
\label{Lambda}
\end{equation}
where ${\rm vec}^2A$ is defined by
\begin{equation}
{\rm vec}^2A =
\begin{bmatrix}
{\rm vec}(A_{--}) & {\rm vec}(A_{-+})\\
{\rm vec}(A_{+-}) & {\rm vec}(A_{++})
\end{bmatrix},
\label{vec1}
\end{equation}
and ${\rm vec}$ is the vectorization operator $({\rm vec}A)_{mL+l, nL+k} = A^{ml}_{nk}$.
When $\Lambda<0$ the chygraph is made of finite size components, while when  $\Lambda>0$ there is a giant component. Since $\Lambda({\rm vec}^2A)=0$ implies $\det({\rm vec}^2A)=0$, we define the pseudo-order parameter
\begin{equation}
\theta = -\det({\rm vec}^2A).
\label{theta}
\end{equation}
In general $\det({\rm vec}^2A)$ has a simpler analytical expression than $\Lambda({\rm vec}^2A)$, making $\theta$ more suitable to determine the percolation threshold when the combinatorial constructions are too complex. However, bear in mind that the sign of $\theta$ cannot be used to determine the percolation phase. For that we're force to calculate $\Lambda({\rm vec}^2A)$.

\subsection{Symbolic calculation}
\label{sec_sym_theta}

The specificities associated with different chygraphs are encoded in the matrices $\langle\kappa\rangle$, $\langle\bar{\kappa}\rangle$, $\langle s\rangle$ and $\langle\bar{s}\rangle$. Once those matrices have been specified the order parameter $\Lambda$ and $\theta$ are calculated using Eqs. (\ref{A})-(\ref{theta}). In general such calculations are cumbersome  and a symbolic calculator is recommended. For example, using symbolic python, we can calculate ${\rm vec2A}(\langle\kappa\rangle, \langle\bar{\kappa}\rangle, \langle s\rangle, \langle\bar{s}\rangle)$ using the python class \verb-vec2A- at \href{https://github.com/av2atgh/chygraph}{github.com/av2atgh/chygraph}. This python class constructs the vectorized matrix ${\rm vec}^2A$ using the vectorization operation $({\rm vec}A)_{mL+l, nL+k} = A^{ml}_{nk}$. Then it calculates $\theta$ and the eigenvalues of ${\rm vec}^2A$ using symbolic python functions. The input parameters are $k$, $K$, $s$ and $S$, corresponding to $\langle\kappa\rangle$, $\langle \bar{\kappa} \rangle$, $\langle s\rangle$ and $\langle\bar{s}\rangle$, respectively.

\section{Graphs and hypergraphs}
\label{sec_graphs_hypergraphs}

The best way to illustrate the applicability of the function ${\rm vec2A}$ is by example. We start with known results for graphs a hypergraphs. Consider a hypergraph with arbitrary vertex degree and hyperedges cardinality distributions. Let $\langle k\rangle$ and $\langle k(k-1)\rangle/\langle k\rangle$ be the average degree and excess nodes degree, respectively.  Let $\langle c\rangle$ and $\langle c(c-1)\rangle/\langle c\rangle$ be the average hyperedges cardinality and excess cardinality, respectively. The hypergraph nodes and hyperedges are said to be present with probability $p$ and $q$, respectively.

A hypergraph is represented by a chygraphs with a layer for nodes and a layer for hyperedges (Fig. \ref{fig_mappings}b). Let us label the nodes by layer 0 and the hyperedges by layer 1. Then
\begin{equation}
\langle\kappa\rangle =
\begin{bmatrix}
0 & p\langle k\rangle\\
0 & 0
\end{bmatrix}
\label{kappa_hypergraph}
\end{equation}
\begin{equation}
\langle\bar{\kappa}\rangle =
\begin{bmatrix}
0 & p \langle\bar{k}\rangle\\
0 & 0
\end{bmatrix},
\label{K_hypergraph}
\end{equation}
\begin{equation}
\langle s\rangle =
\begin{bmatrix}
0 & 0\\
q\langle c\rangle & 0
\end{bmatrix},
\label{s_hypergraph}
\end{equation}
\begin{equation}
\langle\bar{s}\rangle =
\begin{bmatrix}
0 & 0\\
q\langle\bar{c}\rangle & 0
\end{bmatrix}.
\label{S_hypergraph}
\end{equation}
Using {\rm vec2A} we obtain the order parameter (See Supplemental Material at [\href{hhttps://github.com/av2atgh/chygraph/blob/master/supp.ipynb}{github.com/av2atgh/chygraph}], Supp. Eqs. 1-3)
\begin{equation}
\theta_H = pq \langle\bar{k}\rangle \langle\bar{c}\rangle - 1,
\label{theta_hypergraph}
\end{equation}%
\begin{equation}
\Lambda_H = \sqrt{\theta_H+1} - 1.
\label{lambda_hypergraph}
\end{equation}
There is a giant component when $\theta_H\geq0$, in agreement with previous reports \cite{coutinho20,sun21,bianconi23}.

A graph is a restricted version of a hypergraph where the edges contain exactly two nodes. In this case $\langle\kappa\rangle$ and $\langle\bar{\kappa}\rangle$ are given by Eqs. (\ref{kappa_hypergraph}) and (\ref{K_hypergraph}), while
\begin{equation}
\langle s\rangle =
\begin{bmatrix}
0 & 0\\
2q & 0
\end{bmatrix},
\label{s_graph}
\end{equation}
\begin{equation}
\langle\bar{s}\rangle =
\begin{bmatrix}
0 & 0\\
q &  0
\end{bmatrix}.
\label{S_graph}
\end{equation}
Using {\rm vec2A} we obtain the order parameter (See Supplemental Material at [\href{hhttps://github.com/av2atgh/chygraph/blob/master/supp.ipynb}{github.com/av2atgh/chygraph}], Supp. Eqs. 4-6)
\begin{equation}
\theta_G = pq \langle\bar{k}\rangle - 1,
\label{theta_graph}
\end{equation}
\begin{equation}
\Lambda_G = \sqrt{\theta_G+1} - 1,
\label{theta_graph}
\end{equation}
as reported in 1998 by Molloy \& Read \cite{molloy98} for the case $p=q=1$ and in 2000 by Callaway {\em et al} \cite{callaway00} for any $p$ and $q$.

\section{Multiplex}
\label{sec_multiplex}

A multiplex hypergraph with $L$ hyperedge types is mapped to a chygraph with a layer of  atoms representing nodes (layer 0) and $L$ layers representing hypergedge types (Fig. \ref{fig_mappings}c). 

The hypergraph nodes and hyperedges are said to be present with probability $p$ and $q_l$, $l=1\ldots, L$, respectively. I solved the problem of bond percolation ($p=1$) in multiplex networks using methods from multi-type branching processes \cite{vazquez06a}. Later on Allard {\em et al} solved the problem using the more standard generating function formalism for multiplex networks \cite{allard09}. Their generating function calculations contain the precursors to my vectorization technique in Ref. \cite{vazquez22}. Here we recapitulate these results and extend them to the case of multiplex hypergraphs.

For random multiplex hypergraphs $\langle\kappa\rangle$, $\langle\bar{\kappa}\rangle$, $\langle s\rangle$ and $\langle\bar{s}\rangle$ are $(L+1)\times(L+1)$ matrices and the only non-zero matrix elements are
\begin{equation}
\langle\kappa\rangle_{0l} = p\langle k\rangle_l,
\label{kappa_multiplex}
\end{equation}
\begin{equation}
\langle\bar{\kappa}\rangle_{0l} = p\langle\bar{k}\rangle_l,
\label{K_multiplex}
\end{equation}
\begin{equation}
\langle s\rangle_{l0} = q_l\langle c\rangle_l,
\label{s_multiplex}
\end{equation}
\begin{equation}
\langle\bar{s}\rangle_{l0} = q_l\langle\bar{c}\rangle_l,
\label{S_multiplex}
\end{equation}
for $l=1,\ldots,L$.

The cases $L=2$ and $L=3$ are reported in the Supplemental Material at [\href{hhttps://github.com/av2atgh/chygraph/blob/master/supp.ipynb}{github.com/av2atgh/chygraph}], Supp. Eqs. 7-10. For $L=2$ both $\theta$ and $\Lambda$ contain several terms. For $L=3$ the expression of $\theta$ is a very long polynomial and the equation for $\Lambda$ is not worth displaying. It is best to handle these equations with a symbolic calculator and evaluate them numerically for specific parameter sets. 

\subsection{Poisson multiplex}

One may question why $\theta$ gets so complicated. The answer is excess degrees. To illustrate the point, consider a multiplex hypergraph with type dependent Poisson distributions of nodes degrees and hyperedges cardinalities. If $x$ is a random variable with a Poisson distribution then $\langle x(x-1)\rangle / \langle x\rangle =  \langle x\rangle$. Applying this equality for both degrees and cardinalities in Eqs. (\ref{K_multiplex}) and (\ref{S_multiplex}) yields
\begin{equation}
\langle\bar{\kappa}\rangle_{0l} = p\langle k\rangle_l,\ \ \ \ 
\langle\bar{s}\rangle_{l0} = q_l\langle c\rangle_l,
\label{ks_multiplex_poisson}
\end{equation}
for $l=1,\ldots,L$. Using {\rm vec2A} we then obtain  (See Supplemental Material at [\href{hhttps://github.com/av2atgh/chygraph/blob/master/supp.ipynb}{github.com/av2atgh/chygraph}], Supp. Eqs. 11-13)
\begin{equation}
\theta_{MP} = p\sum_{l=1}^L q_l \langle k\rangle_l \langle c\rangle_l - 1,
\label{theta_multiplex_poisson}
\end{equation}
\begin{equation}
\Lambda_{MP} = \sqrt{\theta_{MP}+1} - 1.
\label{Lambda_multiplex_poisson}
\end{equation}
For Poisson multiplex hypergraphs the contribution of hyperedge types becomes additive. This result was previously reported by Sun and Bianconi \cite{sun21} (Eq. 33). Comparing the simplicity of Eq. (\ref{theta_multiplex_poisson}) to that for arbitrary distributions of degrees and cardinalities (See Supplemental Material at [\href{hhttps://github.com/av2atgh/chygraph/blob/master/supp.ipynb}{github.com/av2atgh/chygraph}], Supp. Eqs. 7-10) we arrive to the conclusion that excess averages are the cause of the increased polynomial terms in the expression for $\theta$. 

\begin{figure}[t]
\includegraphics[width=3in]{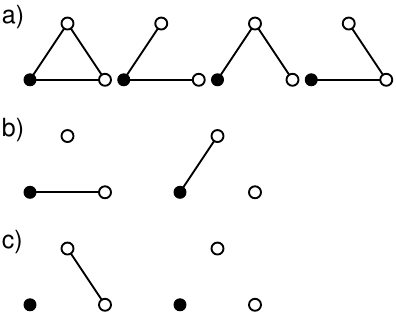}
\caption{Possible intra-complex components for a triangle with links present with probability $q$. The filled circle represents the node from where we arrived to the triangle. a) Excess component size $\bar{s}=2$ with probability $q^3+3q^2(1-q)$. b) Excess component size $\bar{s}=1$ with probability $2q(1-q)^2$. c) Excess component size $\bar{s}=0$ with probability $(1-q)^3+p(1-p)^2$.}
\label{fig_bond_percolation_triangle}
\end{figure}

\subsection{Network motifs}

Real networks contain motifs, subgraphs that are at higher abundance than what expected by chance \cite{watts98, milo02}. If we interpret the node degree as the number of links where the node is included, then we define the type-$l$ degree as the numbers of type-$l$ motifs where the node is included. This interpretation was introduced by Mann {\em et al} to tackle the generating function formalism given the type degree distributions  \cite{mann21}.   The combinatorial construction by Mann {\em et al} can be mapped to a multiplex chygraph, with a layer representing nodes and $L$ additional layers representing each network motif under consideration (links, triangles, ...). Similarly, Bianconi \cite{bianconi21} and Zhao {\em et al} \cite{zhao22} have used simplicial complexes to solve the problem of clique percolation on random graphs, where a click is the fully connected network motif.

The case of network motifs introduces additional complexity with respect to the canonical multiplex hypergraphs discussed above. This complexity is best illustrated by the problem of bond percolation. When links are present with some probability $q$ (bond percolation), then we need to characterize the connectivity within the motifs. This is an example where complexes have fine-grained structure.

I'll expand on this point by solving the problem of bond percolation in a graph with overrepresented triangles. When we reach a triangle via one of its nodes, we can reach different intra-complex component excess sizes (Fig. \ref{fig_bond_percolation_triangle}). Thus, labeling the nodes, links and triangles layers by 0, 1 and 2 respectively we obtain the average and excess average matrices
\begin{equation}
\langle\kappa\rangle = 
\begin{bmatrix}
0 & \langle k\rangle_| &  \langle k\rangle_\triangle\\
0 & 0 & 0\\
0 & 0 & 0
\end{bmatrix}
\label{kappa_triangle}
\end{equation}
\begin{equation}
\langle\bar{\kappa}\rangle = 
\begin{bmatrix}
0 & \langle \bar{k}\rangle_| &  \langle \bar{k}\rangle_\triangle\\
0 & 0 & 0\\
0 & 0 & 0
\end{bmatrix}
\label{K_triangle}
\end{equation}
\begin{equation}
\langle s\rangle = 
\begin{bmatrix}
0 & 0 & 0\\
s_|(q) & 0 & 0\\
s_\triangle(q) & 0 & 0
\end{bmatrix}
\label{s_triangle}
\end{equation}
\begin{equation}
\langle \bar{s}\rangle = 
\begin{bmatrix}
0 & 0 & 0\\
\bar{s}_|(q) & 0 & 0\\
\bar{s}_\triangle(q) & 0 & 0
\end{bmatrix},
\label{S_triangle}
\end{equation}
where
\begin{equation}
s_|(q) = 1+q,
\label{s_|}
\end{equation}
\begin{equation}
\bar{s}_|(q) = q,
\label{S_|}
\end{equation}
\begin{equation}
s_\triangle(q) = 3(q^3+3q^2(1-q)) + \frac{3}{2}3q(1-q)^2 + (1-q)^3,
\label{s_triangle}
\end{equation}
\begin{equation}
\bar{s}_\triangle(q) = 2q(1+q-q^2).
\label{S_triangle}
\end{equation}
Substituting the above matrices into ${\rm vec}^2A$ we obtain the pseudo-order parameter (See Supplemental Material at [\href{hhttps://github.com/av2atgh/chygraph/blob/master/supp.ipynb}{github.com/av2atgh/chygraph}], Supp. Eqs. 14-16)
\begin{eqnarray}
\theta_{NM} &=& \sum_{l=|,\triangle} \bar{s}_l(q) \langle \bar{k}\rangle_l - 1
\nonumber\\
&+& \bar{s}_|(q)\bar{s}_\triangle(q) (
\langle k\rangle_| \langle k\rangle_\triangle
- \langle \bar{k}\rangle_| \langle \bar{k}\rangle_\triangle ),
\label{theta_triangle}
\end{eqnarray}
while the expression for $\Lambda$ is too long to be displayed here.

Let's go over some particular cases to verify this result. When there are no triangles ($\langle \bar{k}\rangle_\triangle = \langle k\rangle_\triangle = 0)$ we recover the order parameter of bond percolation on graphs with arbitrary degree distribution (Eq. (\ref{theta_graph}), $p=1$). When the distributions of nodes participation in links and triangles are Poisson $\langle \bar{k}\rangle_| \langle \bar{k}\rangle_\triangle = \langle k\rangle_| \langle k\rangle_\triangle$ and Eq. (\ref{theta_triangle}) is reduced to (See Supplemental Material at [\href{hhttps://github.com/av2atgh/chygraph/blob/master/supp.ipynb}{github.com/av2atgh/chygraph}], Supp. Eqs. 17-19)
\begin{equation}
\theta_{\rm NMP} = \sum_{l=|,\triangle} \bar{s}_l(q) \langle \bar{k}\rangle_l - 1.
\label{theta_triangle_poisson}
\end{equation}
\begin{equation}
\Lambda_{NMP} = \sqrt{\theta_{NMP}+1} - 1.
\label{Lambda_multiplex_poisson}
\end{equation}%
For Poisson distributions the contribution of links and triangles is additive, as it is for multiplex hypergraphs with Poisson distributions Eq. (\ref{theta_multiplex_poisson}).

\begin{figure}[t]
\includegraphics[width=3in]{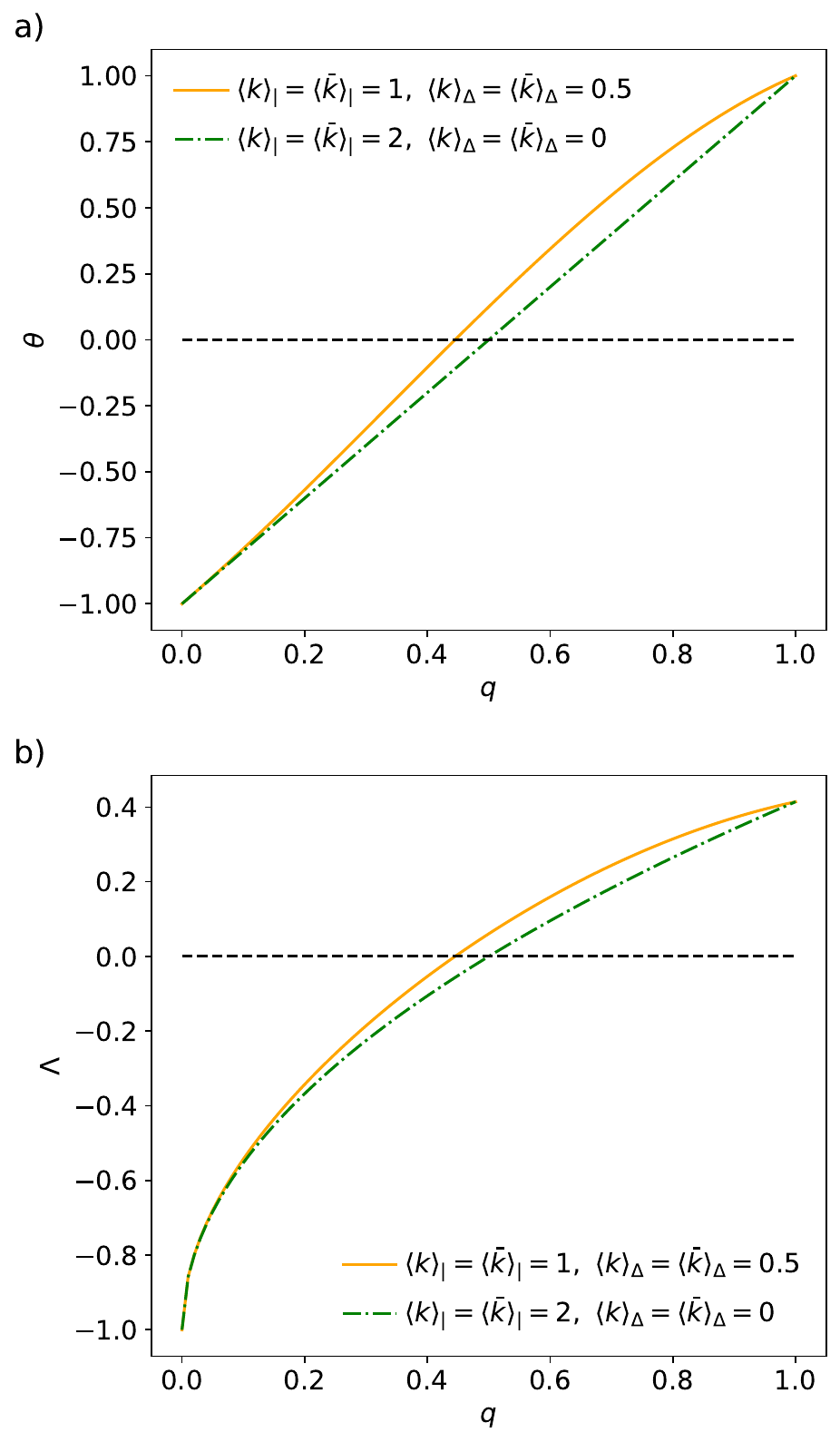}
\caption{The $q$ dependency of a) the pseudo-order parameter $\theta$ and b) the order parameter $\Lambda$ for the case of two Poisson networks with the same number of links. One with triangles (solid line) and one with no triangles (dashed-dotted line).}
\label{fig_theta_triangle}
\end{figure}

An interesting question is when is the network more resilient, when links are distributed independently of or within triangles, provided the total number of links is the same. The number of links independent of and within triangles are $E_| = \langle k\rangle_|/2$ and $E_| = 3E_\triangle = \langle k\rangle_\triangle$, respectively. To preserve the total number of links, the Poisson network with parameters  $(\langle k\rangle_|, \langle k\rangle_\triangle)$ should be compared to the network without triangles $(\langle k\rangle_| + 2\langle k\rangle_\triangle, 0)$. Using Eq. (\ref{theta_triangle_poisson}) we obtain
\begin{equation}
\theta_{\rm P}(\langle k\rangle_|, \langle k\rangle_\triangle) -
\theta_{\rm P}(\langle k\rangle_| + 2\langle k\rangle_\triangle, 0) = 
2q^2(1-q)\langle k\rangle_\triangle.
\label{delta_theta}
\end{equation}
This difference is greater than 0 for $q>0$, as it is illustrated in Fig. \ref{fig_theta_triangle} for a specific set of parameters. With a fixed number of links, random networks where links are distributed as triangles are more robust than those where links are not distributed as triangles. Since spreading of an infectious agent in a network is equivalent to bond percolation \cite{newman02}, the latter means that having links linked to triangles increases the propensity for spreading. This coincides with the earlier observation by Newman \cite{newman03}.

\section{Interacting graphs and hypergraphs}
\label{sec_interacting}

Interacting networks are layers of different type networks connected by inter-network links. Interacting hypergraphs are layers of different type hypergraphs connected by inter-hypergraphs hyperedges. Figure \ref{fig_interacting} shows the mapping of two interacting hypergraphs to a chygraph with 5 layers. Two layers of nodes and hyperedges for each hypergraphs, plus one layer of hyperedges containing nodes from both hypergraphs.

For two interacting hypergraphs we can assign labels from 0 to 4 to the different layers in Fig. \ref{fig_interacting}. For $g\gg1$ interacting hypergraphs we better use some clever indexing. Here I restrict my attention two hyperedges containing nodes of at most two distinct hypergraphs. I choose to label the hypergraphs $H_l, l=0,\ldots,g-1$ and assign their nodes to layers $l=0,\ldots,g-1$, respectively. The hyperedges containing nodes from $H_l$ and $H_m$, $l\leq m$, are assigned to the layer indexed by
\begin{eqnarray}
i_{lm} &=& g + \sum_{m=0}^{l-1}(g-m) + m-l
\nonumber\\
&=& g + \frac{2g - (l-1)}{2}l + m-l
\label{ndexing_interacting}
\end{eqnarray}

For the case of two interacting hypergraphs, the hyperedges of $H_0$ are assigned to layer $i_{00} = 2$, those of $H_1$ to $i_{11} = 4$ and the interacting hyperedges to $i_{01} = 3$. With this notation we obtain the mean and excess mean matrices of 
\begin{equation}
\langle\kappa\rangle =
\begin{bmatrix}
0 & 0 & \langle k\rangle_{02} & \langle k\rangle_{03} & 0\\
0 & 0 & 0 & \langle k\rangle_{13}  & \langle k\rangle_{14}\\
0 & 0 & 0 & 0 & 0\\
0 & 0 & 0 & 0 & 0\\
0 & 0 & 0 & 0 & 0
\end{bmatrix}
\label{kappa_interacting}
\end{equation}
\begin{equation}
\langle\bar{\kappa}\rangle =
\begin{bmatrix}
0 & 0 & \langle \bar{k}\rangle_{02} & \langle \bar{k}\rangle_{03} & 0\\
0 & 0 & 0 & \langle \bar{k}\rangle_{13}  & \langle \bar{k}\rangle_{14}\\
0 & 0 & 0 & 0 & 0\\
0 & 0 & 0 & 0 & 0\\
0 & 0 & 0 & 0 & 0
\end{bmatrix}
\label{K_interacting}
\end{equation}
\begin{equation}
\langle s \rangle =
\begin{bmatrix}
0 & 0 & 0 & 0 & 0\\
0 & 0 & 0 & 0 & 0\\
\langle c\rangle_{20} & 0 &  0 & 0 & 0\\
\langle c\rangle_{30} & \langle c\rangle_{31} &  0 & 0 & 0\\
0 & \langle c\rangle_{41} &  0 & 0 & 0
\end{bmatrix}
\label{s_interacting}
\end{equation}
\begin{equation}
\langle \bar{s} \rangle =
\begin{bmatrix}
0 & 0 & 0 & 0 & 0\\
0 & 0 & 0 & 0 & 0\\
\langle \bar{c}\rangle_{20} & 0 &  0 & 0 & 0\\
\langle \bar{c}\rangle_{30} & \langle \bar{c}\rangle_{31} &  0 & 0 & 0\\
0 & \langle \bar{c}\rangle_{41} &  0 & 0 & 0
\end{bmatrix}
\label{bar_s_interacting}
\end{equation}

We can go ahead and plug in these matrices in ${\rm vec2A}$ to obtain $\theta$ and $\Lambda$(See Supplemental Material at [\href{hhttps://github.com/av2atgh/chygraph/blob/master/supp.ipynb}{github.com/av2atgh/chygraph}], Supp. Eqs. 20-22). For the general case we obtain a polynomial with several terms.

\subsection{Interacting Poisson hypergraphs and graphs}

For hypergraphs with Poisson distributions of degrees and cardinalities $\langle\bar{\kappa}\rangle = \langle\kappa\rangle$,   $\langle\bar{s}\rangle = \langle s\rangle$. Theequations for $\theta$ and $\Lambda$ are simplified (See Supplemental Material at [\href{hhttps://github.com/av2atgh/chygraph/blob/master/supp.ipynb}{github.com/av2atgh/chygraph}], Supp. Eqs. 23-25), but they are still too long to display here.

For two interacting graphs with Poisson degree distributions the average and excess average matrices are
\begin{equation}
\langle\kappa\rangle =
\begin{bmatrix}
0 & 0  &\langle k\rangle_{02} & \langle k\rangle_{03} & 0\\
0 & 0 & 0 & \langle k\rangle_{13}  & \langle k\rangle_{14}\\
0 & 0 & 0 & 0 & 0\\
0 & 0 & 0 & 0 & 0\\
0 & 0 & 0 & 0 & 0
\end{bmatrix},
\label{kappa_interacting_graphs}
\end{equation}
\begin{equation}
\langle\bar{\kappa}\rangle = \langle\kappa\rangle,
\label{K_interacting_graphs}
\end{equation}
\begin{equation}
\langle s \rangle =
\begin{bmatrix}
0 & 0 & 0 & 0 & 0\\
0 & 0 & 0 & 0 & 0\\
2 & 0 &  0 & 0 & 0\\
1 & 1 &  0 & 0 & 0\\
0 & 2 &  0 & 0 & 0
\end{bmatrix},
\label{s_interacting_graphs}
\end{equation}
\begin{equation}
\langle \bar{s} \rangle =
\begin{bmatrix}
0 & 0 & 0 & 0 & 0\\
0 & 0 & 0 & 0 & 0\\
1 & 0 &  0 & 0 & 0\\
0 &  0 & 0 & 0 & 0\\
0 & 1 &  0 & 0 & 0
\end{bmatrix}.
\label{bar_s_interacting_graphs}
\end{equation}
Notice that $\langle s\rangle_{40} = \langle s\rangle_{40} = 1$. The complexes at layer 4, representing the inter-graph links, include only one node from each of the two interacting graphs. Furthermore, $\langle \bar{s}\rangle_{40} = \langle \bar{s}\rangle_{40} = 0$. That means, if you come from one graph using the intercating links you cannot go back to the same graph. Plugging in Eqs. (\ref{kappa_interacting_graphs})-(\ref{bar_s_interacting_graphs}) we obtain (See Supplemental Material at [\href{hhttps://github.com/av2atgh/chygraph/blob/master/supp.ipynb}{github.com/av2atgh/chygraph}], Supp. Eqs. 23-25)
\begin{equation}
\theta_{IGP} = \langle k\rangle_{04}\langle k\rangle_{24} 
- (\langle k\rangle_{01} - 1)(\langle k\rangle_{23} - 1),
\label{theta_interacting_graphs}
\end{equation}
\begin{eqnarray}
\Lambda_{IGP} &=& \frac{1}{\sqrt{2}} \biggl(
\langle k\rangle_{02} + \langle k\rangle_{14} 
\nonumber\\
&+&
\sqrt{(\langle k\rangle_{02}-\langle k\rangle_{14})^2  + 
4\langle k\rangle_{03}\langle k\rangle_{13} } \biggr)^{1/2}
\nonumber\\
&-& 1.
\label{Lambda_interacting_graphs}
\end{eqnarray}
The result for $\theta_{IGP}$ was previously obtained by Leicht and Souza \cite{leicht09}. You can crosscheck that $\Lambda_{IGP}=0$ is equivalent to $\theta_{IGP}=0$.

The extension to interacting graphs with arbitrary degree distributions and any number of layers is straightforward with the symbolic calculator, albeit the results are very long equations. The case of two interacting graphs with arbitrary degree distributions is reported in the Supplemental Material at [\href{hhttps://github.com/av2atgh/chygraph/blob/master/supp.ipynb}{github.com/av2atgh/chygraph}], Supp. Eqs. 23-25. This result recapitulates the calculations of Bianconi for percolation in multilayer interacting graphs \cite{bianconi17}. 

\section{Conclusions}
\label{sec_conclusions}

Chygraphs are a general combinatorial construction that includes as special cases many of the higher order structures defined so far. When that mapping is possible, we can use the generating function calculation for the order parameter characterizing the  emergence of a giant component in chygraphs. This is done by the symbolic calculation reported in this work. 

When the degree of participation in complexes follow Poisson distributions there is some simplification in the analytical expressions. In many instances $\Lambda = \sqrt{\theta} - 1$, where $\Lambda$ and $\theta$ are the largest eigenvalue and the determinant of the generalized average excess component tensor $-A$. However, there are exceptions, as shown for the case of two interacting graphs with Poisson degree distributions.

The example of bond percolation in graphs with overrepresented triangles illustrates the relevance of having complexes with a fine-structure. It helps to decompose the problem into the large scale connectivity and the local connectivity within the complex.

The recapitulation of previous results demonstrates the range of applicability of chygraphs and the convenience of the symbolic calculations. When we tackle a problem in chygraphs we're tackling the  problem for a wide range of higher order networks. This is the way to move forward.

Future work should address the applicability chygraphs percolation in real scenarios. A new challenge is the problem of interacting hypergraphs. Ecology is a good example, with organisms acting as hyperedges of enzymes, food acting as hyperedges of metabolites and the the ability of enzymes to process certain metabolites as interacting hyperedges. The symbolic calculations introduced will help researchers to investigate the percolation properties of these systems without the need of cumbersome and error-prone calculations.

\bibliographystyle{apsrev4-1}


%

\end{document}